# The Origin and Value of Disagreement Among Data Labelers: A Case Study of Individual Differences in Hate Speech Annotation

Yisi Sang and Jeffrey Stanton

Syracuse University, Syracuse NY 13244, USA
yisang@syr.edu, jmstanto@syr.edu

**Abstract.** Human annotated data is the cornerstone of today's artificial intelligence efforts, yet data labeling processes can be complicated and expensive, especially when human labelers disagree with each other. The current work practice is to use majority-voted labels to overrule the disagreement. However, in the subjective data labeling tasks such as hate speech annotation, disagreement among individual labelers can be difficult to resolve. In this paper, we explored why such disagreements occur using a mixed-method approach – including interviews with experts, concept mapping exercises, and self-reporting items – to develop a multidimensional scale for distilling the process of how annotators label a hate speech corpus. We tested this scale with 170 annotators in a hate speech annotation task. Results showed that our scale can reveal facets of individual differences among annotators (e.g., age, personality, etc.), and these facets' relationships to an annotator's final label decision of an instance. We suggest that this work contributes to the understanding of how humans annotate data. The proposed scale can potentially improve the value of the currently discarded minority-vote labels.

**Keywords: data labeler, label, disagreement, hate speech, multidimensional scale, content moderation**

## 1 Introduction

With currently predominant machine learning techniques, humans support the training of machine learning algorithms by providing annotation data [36], often in the form of binary categorizations. Such data work well for training perceptual-level classifiers – for example, those used for labeling a photograph as a cat or dog. In contrast, consider a more complex "social computing" application, where we want to develop a system that can characterize the toxicity of a social media post. Taking into account the messiness of social reality, people's perceptions of toxicity vary substantially. Their labels of toxic content are based on their own perceptions. We refer to the first type of data labeling tasks as "objective data labeling", and the second type of tasks as "subjective data labeling", and in this paper, we are interested in the labelers' disagreements for subjective data labeling tasks.

Subjectiveness usually arises from what some researchers refer to as "latent content". The idea of latent content refers to the presence in text of meaning that must be inferred because it is not manifest in the dictionary meanings of individual words. In an influential paper about the validity of content analysis, Potter and Levine-Donnerstein [42] distinguished latent content into pattern and projective subtypes. "Pattern" latent content can be inferred by recognizing a meaningful pattern across recurrent cues present in the text. The detection of pattern latent content depends on a judge's recognition of key configurations of elements in the material. As an example, imagine a statement that on its face seems to indicate one thing, but the subsequent appearance of a sarcasm indicator reverses that meaning. In contrast, "projective" latent content may require a more complex approach. Key elements in projective latent content are symbols that "require viewers to access their pre-existing mental schema in order to judge the meaning in the content" [42]. This point highlights the fact that different annotators may bring different mental schemas to the stimulus evaluation process and thereby may also produce differing judgments about the same content.

The present study is concerned with the annotation of latent content in hate speech, where the mental schemas that human annotators bring to the annotation process may result in systematic variation in judgments about content. Psychometrically sound measurement instruments are generally construed to reflect the underlying dimensional structure of the unobservable concepts they measure. A psychometrically sound instrument that captures the dimensionality of judgments about a set of stimuli (i.e., examples of hate speech) may thus reveal some aspects of how annotators conceptualize content and may also offer analytical insights about variability in their judgments about the content. Therefore, in this study we first developed a multidimensional scale to capture hate speech judgments using standard psychometric scale development methods. Then we conducted multigroup confirmatory factor analysis on n=170 annotators' evaluations of hate speech. Our analysis focused on addressing these research questions:
Research Question 1: What are some basic judgment dimensions for human annotations of hate speech?
Research Question 2: How does the age of a human judge relate to the annotation of hate speech?
Research Question 3: How do personality factors relate to the annotation of hate speech?

## 2 Related Work

### 2.1 Definitions and Elements of Hate Speech

Legal frameworks, social science research, and social media companies have all generated definitions of hate speech, with their particular goals driving the definitional approach they take [50]. Legal frameworks for hate speech are influenced by the social values of speech regulation in particular national contexts [41]. For example, Article 20 of the International Covenant on Civil and Political Rights (ICCPR) states that "[a]ny advocacy of national, racial or religious hatred that constitutes incitement to discrimination, hostility or violence shall be prohibited by law." In 2016, to counter online hate speech, the commission of EU developed a "code of conduct" agreement with Facebook, Microsoft, Twitter and YouTube. The government of China has established similar mechanisms to counter online hate speech as a part of its Internet censorship policy [53].

Social media platforms have developed their own definitions of hate speech. These definitions guide their content moderation strategies. For instance, Facebook defines hate speech as "a direct attack on people based on what we call protected characteristics – race, ethnicity, national origin, religious affiliation, sexual orientation, caste, sex, gender, gender identity, and serious disease or disability" [60]. Twitter has a hateful conduct policy using the definition of promoting violence, "against or directly attacking, or threatening other people on the basis of race, ethnicity, national origin, caste, sexual orientation, gender, gender identity, religious affiliation, age, disability, or serious disease" [61].

One early academic effort to define hate speech was Richard Delgado's article "Words that Wound" [17]. Based on Delgado's work, Matsuda argued that a narrow class of racial hate speech that causes serious harm should be considered as a crime [30][31]. Moran and Ward emphasized the intention of hate speech. Moran defined hate speech as "speech that is intended to promote hatred against traditionally disadvantaged groups"[35]. Marwick and Miller [29] proposed three distinct elements that were used in defining hate speech: 1) a content-based element; 2) an intent-based element; and 3) a harms-based element [29]. Benesch [6] defined five aspects of dangerous hate speech: the speaker, the audience, the speech act itself, the social and historical context, and the means of dissemination. Parekh emphasized the target of hate speech [40]. Sellars[50] proposed a framework to conceptualize hate speech including considering the target, content, harm, speaker, result, and context of verbal hate speech.

In summary, our literature review suggests that hate speech judgments re multidimensional in nature. Previous work has considered the status of the source, the status of the target – including whether the target is a (member of a) protected class, the intent of the speech, and the impacts of the speech in the world. While this is probably not an exhaustive list of the aspects or dimensions of hate speech, these perspectives clearly indicate that a binary approach to detecting, flagging, or moderating hate speech oversimplifies the underlying judgments that people make about toxic content.

### 2.2 Annotation of Hate Speech

Hat speech could be construed as an umbrella term that includes many kinds of problematic user-generated content [49]: abusive language [38][54], cyberbullying [52][1], offensive language [11][58], misogynist language [22][18], toxic language [34][23], threatening language [39], aggression [26], and hate speech [55][24]. In various research communities, these terms are sometimes considered distinctive and sometimes considered as overlapping. Many previous annotation efforts have used simple binary classification decisions to annotate content in any one of these categories or to designate a choice among a small number of mutually exclusive categories. For example, in Chen et al. [11], YouTube comments were labeled based on whether they are offensive or not. Some researchers have constructed datasets that distinguish facets in various expressions of online hate speech. For example, in Founta [19], annotators were required to label tweets as offensive, abusive, hateful speech, aggressive, cyberbullying, spam, and normal. Some researchers have devised hierarchical annotation schemas that require annotators to code texts on nested levels. For example, in Basile et al. [4], tweets were first coded as hateful or non-hateful. Hateful tweets were further coded based on 1) whether the target was a generic group of people or a specific individual; 2) whether the tweet was aggressive or not. In applying annotation schemes to hate speech, much research has assumed that every instance can be placed into two or more of these mutually exclusive categories. As an alternative, Sanguinetti et al. [48] designed an annotation method that considered factors contributing to the definition of hate speech. Assimakopoulos et al. [2]

considered the polarity of hate speech and annotated hate speech based on discursive strategies including who was the attitude target and how the attitude was expressed.

Among the annotation approaches described above, two popular methods for assessing quality are inter-rater agreement and gold standard corpus test, which assume that there is a single correct truth with respect to assignment of speech examples to particular annotation classes. However, given the complex dimensionality of hate speech, it may be erroneous to assume that there is only one universally acceptable annotation choice for a given stimulus. People's perceptions of hate speech are influenced by many factors. Previous research has found that gender [32], ethnicity [14], racial attitudes [56] [47], value placed on freedom of speech [15], context [14], target [27], empathy [13], ways of knowing [13] [57], implicitness of the hate speech [7], and relationality [13] may all influence people's perception of hate speech. For example, Roussos and Dovidio found that for Black-targeted hate speech, people higher in anti-Black prejudice tended to invoke freedom of speech rights and perceived it less as a hate crime. When the hate target was Black versus White, people with low-prejudice rated the act as less protected by freedom of speech rights and more strongly as a hate crime [47]. Cowan and Hodge found that in the public setting, when an argumentative response occurred, speech was perceived as more offensive than when no response occurred. In a private setting, no response by the target led to higher offensiveness rating than when the target responded [14]. Cowan and Cowan illustrated that the perceived harm of hate speech was positively related to empathy [13].

Both the proliferation of alternative conceptualizations and the widespread use of simplistic annotation methods are arguably hindering the progress of research on online hate speech. For example, Ross, et al. [45] found that showing Twitter's definition of hateful conduct to annotators caused them to partially align their own opinions with the definition. This realignment resulted in very low interrater reliability of the annotations (Krippendorff's $\alpha = 0.29$). Davidson et al.[16] found racial bias in hate speech and abusive language detection datasets. Classifiers trained on these datasets tend to predict tweets written in African-American English vernacular as abusive at substantially higher rates than expected. Awal et al. [3] found that there was annotation inconsistency in popular datasets that are widely used in online hate speech detection. Even expert content moderators often disagree with each other: A Fleiss' kappa for comments annotated by two moderators was only 0.46 [28].

In summary, many alternative annotation methods exist for hate speech. Most of these take the approach of assuming that all speech examples can be assigned to mutually exclusive categories. Gold standard databases take this assumption to its logical conclusion by converging opinions of experts on a set of class assignments assumed to be immutable. While some research considered the possibility that characteristics of hate speech may exist on a set of continua [25], little work has explored the possibility that characteristic differences among human evaluators may influence annotation decisions – whether those decisions include class assignment, ordered ratings, or both. In this light, developing a deeper understanding of how differences among hate speech evaluators cause variation in class assignments or ratings could be valuable in creating and deploying hate speech detection and mitigation systems.

## 3 Method

### 3.1 Development of a Multidimensional Scale for Evaluating Hate Speech

We planned to use confirmatory factor analysis on a set of newly created scale items in an effort to establish some basic evaluative dimensions of hate speech. In the initial work towards this goal, we followed Netemeyer et al.'s 4-step scale creation process [37].

**Construct Definition and Content Domain**: In this step, we identified the facets and domains of the constructs to be assessed. We began by conducting a literature review on the definitions of hate speech and collated 98 definitions from academic papers, law, and online platforms. From the collected material, we compiled a raw list of distinctive elements of online hate speech. To understand people's understanding of online hate speech, we conducted seven semi-structured interviews in October 2020 with researchers who experienced online hate speech or whose research is related to hate speech. We started with questions such as "What do you consider to be hate speech?" Then we examined how participants evaluate hate speech including what factors make them feel that a certain speech instance is hateful. We also asked participants their reactions to hate speech they encountered. Each interview lasted approximately 30 minutes. An inductive coding method [8] was conducted to document the themes of people's evaluations of hate speech. We matched these coded themes to elements from the literature. If the theme of the interview did not appear in the literature, it was treated as a new element. Then we presented each theme to ten hate speech researchers on index cards using an online research platform Optimal Workshop [62]. The experts sorted these cards into categories of their own individual devising. By collapsing across the many similarities among these expert-generated categories, we developed an eight-facet taxonomy of hate speech: Characteristics of the Speaker, Characteristics of the Target, Content of the Speech, Context of the Speech, Impact of the Speech, Intentions of the Speech, Language Style, and Miscellaneous Identification Cues. Table 1 depicts the eight categories.

Table 1. Interview Results: Eight-facets of Hate Speech

| Factor | Description |
| --- | --- |
| Characteristics of the Speaker | Hate speech initiator's attributes |
| Characteristics of the Target | Hate speech recipient/victim's attributes |
| Content | Whether the speech instance expressed violence tendencies, insults to human dignity, slander, moral violation, radicalization, showing fear, and/or groundless accusation. |
| Context | The circumstances where the speech is conducted including cultural, social, political contexts and the condition of the circumstance i.e., private, or public. |
| Impact of the Speech | The outcomes of the speech instance on the targeted individual or on individuals in general |
| Intention of the Speech | The idea or plan of posting the speech instance |
| Language Style | The language style of the speech instance including vocabulary/word choice, sentence structure, figure of speech, whether it is dialogue, monologue, or reported speech etc. |
| Miscellaneous Identification Cues | Catch-all category for other elements of the speech instance |

**Generating and Judging Candidate Items**: In the present study, we focused on Content of the Speech, Characteristics of the Target, and Impact of the Speech in item generation. We used the subcategories of Content of the Speech (i.e., violence tendencies, insult to human dignity, slander, moral violation, radicalization, showing fear, and groundless accusation). We generated and had experts evaluate an initial pool of items to reflect the constructs and ensured that the items were understandable by prospective respondents. The 10 experts who familiar with the elements were tapped again to create items based on the definitions of constructs. Then the experts evaluated these items based on the construct definition and content domain, i.e., whether the item was representative of the construct and the wording of item is understandable. This step resulted in 39 proposed items.

**Developing and Refining the Scale**: In this step, we fielded a survey with one hate speech example and one neutral (non-hateful speech) example each harvested from a popular social media platform. We collected two rounds of data (n=50 and n=120) with slight wording adjustments to the survey between the two rounds. Most respondents were recruited from the research platform Prolific.co [63], where we ensured a minimum pay rate of $9.60 USD per hour. Each survey was completed in less than 10 minutes. Table 2 shows the constructs and items. The median age group of participants was 18-33. 96 participants were male, 69 participants were female, three participants identified themselves as non-binary, and two participants preferred not to report their gender.

**Analyzing the Scale**: In this step, we used Exploratory Factor Analysis (EFA) to identify possible underlying factor structures and refined the item set to its most efficient form, while maintaining acceptable reliability. Then we used Confirmatory Factor Analysis (CFA) to assess the previously identified underlying factor structure.

Table 2. Constructs and Items

| Constructs | Description | List of items |
| --- | --- | --- |
| Violence tendency | Violence tendency refers to the verbal attack, the commitment, threat, or incitement of violence [6]. | 1. This post encourages violence. <br> 2. This post threatens violence toward a specific group. <br> 3. This post attacks someone. <br> 4. This post could cause a public safety problem. <br> 5. This post is a threat of an attack. |
| Impact | Impact of the speech describes the influence and outcomes of the speech such as distorting a perception of a fact, stirring up social panic, delivering conspiracy, and causing social secession. | 6. This post probably distorts the facts. <br> 7. This post uses exaggeration to stir up trouble. |

| Constructs | Description | List of items |
|---|---|---|
| | The impact can be on the individual or on the society. | 8. This post refers to rumors that are untrue. |
| | | 9. This post tries to divide people into a good group and an evil group. |
| | | 10. This post injures the status of the target group. |
| | | 11. This post results in mental distress to the target. |
| Insult to human dignity | Insult to human dignity describes the insult of an individual's humanity such as degrading/judging/discriminating/denigrating people based on their identity, vulnerability, or stereotypes. | 12. The post degrades people based on who they are. |
| | | 13. The post judges people based on stereotypes. |
| | | 14. This post discriminates against vulnerable people. |
| | | 15. This post describes other people as inferior. |
| | | 16. This post criticizes people unfairly. |
| Slander | Slanderous remark refers to the false spoken statement damaging to a person's reputation. | 17. This post insults other people. |
| | | 18. This post tries to damage someone's reputation. |
| | | 19. This post criticizes others falsely. |
| | | 20. This post paints others with a broad brush. |
| Moral violation | Moral violation refers the violation of moral standards such as fairness, respect, care, and sanctity. | 21. This post shows bigotry. |
| | | 22. This post shows intolerance. |
| | | 23. This post shows a lack of respect for others. |
| Radicalization | Radicalization refers to the extreme political, social, or religious ideals and aspirations shown in the speech. | 24. This writer dislikes minorities. |
| | | 25. This writer takes an uncivilized viewpoint. |
| | | 26. This writer probably believes that one group is superior and another group is inferior. |
| Showing Fear | Showing fear refers to expressing fear towards a specific population such as misogyny, homophobia, and xenophobia. | 27. This writer probably dislikes transgender people. |
| | | 28. This writer probably fears foreigners. |
| | | 29. This writer probably dislikes gay men. |

| Constructs | Description | List of items |
|---|---|---|
| Target | The characteristics of the targets focuses on the personality and features of the receivers and victims of the speech [59]. | 30. This writer probably fears women.<br>31. The writer of this post probably feels powerless.<br>32. This post targets a large number of people.<br>33. This post targets a specific person.<br>34. This post targets people based on who they are.<br>35. This post targets people based on their ethnic background. |
| Groundless accusation | Groundless accusation refers to a claim or allegation of wrongdoing that is untrue and/or otherwise unsupported by facts. | 36. The target of this post has done nothing to provoke this attack.<br>37. The target of this post cannot address the writer's problem.<br>38. The writer of this post believes that the target cannot change.<br>39. The target of this post is powerless to address the writer's concern. |

**3.2 Multiple Group Confirmatory Factor Analysis**

Multiple group analysis is a common method to assess group differences, which amounts to doing separate latent variable models of some kind across the groups of interest. We used multiple group analysis to uncover the influence of age and personality on the evaluation of hate speech.

When collecting participants' evaluations of hate speech using the multidimensional scale, we also asked them to complete the big five personality inventory (BFI-10) [44]. The Big Five Inventory (BFI) is a widely used self-reported personality measurement that has five dimensions: Extraversion is characterized by sociability, talkativeness, excitability, and lots of emotional expression [43]; Agreeableness includes traits like sympathetic, kind, and affectionate; Conscientiousness is featured by high levels of thoughtfulness, good impulse control, and goal-directed behaviors [43]; Neuroticism include tense, moody, and anxious; Openness is characterized by having wide interests and being imaginative and insightful. We used the BFI-10 to reduce the length of the questionnaire. Completion time for self-report surveys is negatively correlated with initial willingness to participate [20] and low response rates can adversely affect reliability and validity [44].

To understand the differences of evaluations of hate speech, we analyzed measurement invariance in latent variables. We grouped respondents by personality profile and estimated separate models using a typical sequence for assessing measurement invariance: configural, weak, strong, and strict invariance [5]. Configural invariance is the most basic level, and just indicates that the latent variable model has the same structure in all the groups. Weak invariance constrains loadings to be equal between groups for a given indicator. Latent (co)variances are allowed to vary among groups. In the measurement of strong invariance, all intercepts are constrained to be equal among groups but can vary within a group. Latent means may change among groups. The measurement of strict invariance constrains the error variances to be the same across groups.

## 4 Result

**4.1 Exploratory Factor Analysis**

The scale building process concludes with weeding out poorly functioning items to establish a subset of items that addresses each subscale dimension with an acceptable level of reliability. As a diagnostic, we first conducted exploratory factor analysis (EFA) for each construct. The Cronbach's alpha based upon the covariances of all subscales were above 0.80 except for Target, which had an alpha of 0.66. Item statistics showed that after deleting item 2 the Cronbach alpha of Target increased to 0.73. Cronbach alpha above 0.7 is considered acceptable [12] in scales under development. A parallel analysis of eigenvalues suggested that either a four-factor or a five-factor solution was the best fit for the data. The results of the five-factor solution appear in Table 3.

Table 3. Five-Factor EFA Results

| Items | factor 1 | 2 | 3 | 4 | 5 | Dimension |
|---|---|---|---|---|---|---|
| 1 | **0.796** | | | | | |
| 2 | **0.911** | | | | | |
| 3 | **0.676** | | | | 0.240 | Violence tendency |
| 4 | **0.922** | | | | | |
| 5 | **0.918** | | | | | |
| 6 | | | **0.538** | 0.202 | | |
| 7 | 0.225 | -0.105 | **0.585** | | 0.141 | |
| 8 | 0.191 | | **0.620** | | -0.161 | Impact |
| 12 | **0.412** | 0.113 | 0.315 | | 0.197 | |
| 13 | | | **0.618** | 0.104 | 0.177 | |
| 14 | 0.324 | 0.227 | **0.405** | | | |
| 15 | 0.286 | 0.258 | 0.199 | 0.117 | 0.296 | Insult to human dignity |
| 16 | 0.112 | 0.217 | **0.642** | | -0.135 | |
| 17 | 0.298 | 0.162 | 0.268 | | 0.328 | |
| 18 | | | **0.713** | | | Slander |
| 19 | 0.111 | 0.221 | **0.464** | | 0.135 | |
| 20 | | 0.208 | 0.388 | 0.115 | 0.272 | |
| 22 | 0.145 | 0.180 | 0.266 | 0.116 | **0.461** | Moral violation |
| 23 | 0.223 | 0.118 | 0.236 | 0.121 | **0.486** | |
| 24 | 0.343 | 0.326 | | | 0.260 | |
| 25 | **0.428** | 0.261 | 0.109 | | 0.303 | Radicalization |
| 26 | 0.304 | 0.275 | 0.135 | 0.101 | 0.355 | |
| 27 | 0.161 | **0.657** | | | 0.153 | |
| 28 | 0.194 | **0.523** | | 0.160 | 0.213 | |
| 29 | | **0.921** | | | | Showing Fear |
| 30 | | **0.740** | | 0.109 | | |
| 31 | | **0.551** | | 0.253 | | |

| Item | factor 1 | 2 | 3 | 4 | 5 | Dimension |
|---|---|---|---|---|---|---|
| 32 | 0.211 | 0.111 | | | 0.169 | |
| 34 | 0.236 | 0.165 | 0.172 | | 0.387 | Target |
| 35 | | **0.617** | 0.284 | | -0.134 | |
| 36 | 0.255 | 0.157 | -0.130 | **0.634** | -0.261 | Groundless accusation |
| 37 | | | | **0.836** | | |
| 39 | | | 0.109 | **0.777** | | |

The EFA suggested a sufficient set of items for these five factors: (a) Violence tendency - This factor has an eigenvalue of 4.889 and accounts for 14.8% of the variance; (b) Showing fear - This factor has an eigenvalue of 3.474 and accounts for 10.5% of the variance; (c) The third factor is measured by items from Impact, Insult to human dignity, and Slander. This factor has an eigenvalue of 3.403 and accounts for 10.3% of the variance. (d) Groundless accusation - This factor has an eigenvalue of 2.007 and accounts for 6.1% of the variance; (e) Moral violation - This factor has an eigenvalue of 1.596 and accounted for 4.8% of the variance. Items 15, 17, 20, 24, 26, 32, and 34 were dropped from subsequent analysis based on a poor pattern of loadings.

## 4.2 Confirmatory Factor Analysis of the Five-Factor Model

We fit a five-factor confirmatory model using lavaan [46] in R. We standardized the latent factors, allowing free estimation of all factor loadings and used maximum likelihood to estimate the model.

We started the model fitting with the Violence Tendency latent variable. Modification indices suggested that item 5, item 12, and item 25 had excess correlations to other items. After dropping these, CFI increased from 0.94 to 0.99 and RSMEA dropped from 0.194 to 0.037. Then we added the Showing fear latent variable. Modification indices indicated that the model fit would improve by removing item 29, item 31, and item 35. We removed these items and added the Moral violation latent variable. The results indicated that item 38 should be removed. After removing item 38, RMSEA decreased from 0.067 to 0.04 and CFI increased from 0.978 to 0.994. Next, we added Groundless Accusation. The results showed that item 38 should be dropped. After dropping item 38 the RMSEA decreased from 0.067 to 0.040 and the CFI increased from 0.978 to 0.994. Finally, items 22 and 23 were retained as indicators of Impact. This modification improved CFI from 0.945 to 0.991 and decreased RMSEA from 0.082 to 0.042. For the final model, fit statistics were as follows: CFI=.991, RMSEA=.042, SRMR=.032, BIC= 6054.32. These are generally considered acceptable levels for CFA models, except for BIC, which is not a standardized measure. Figure 1 visualizes the model.

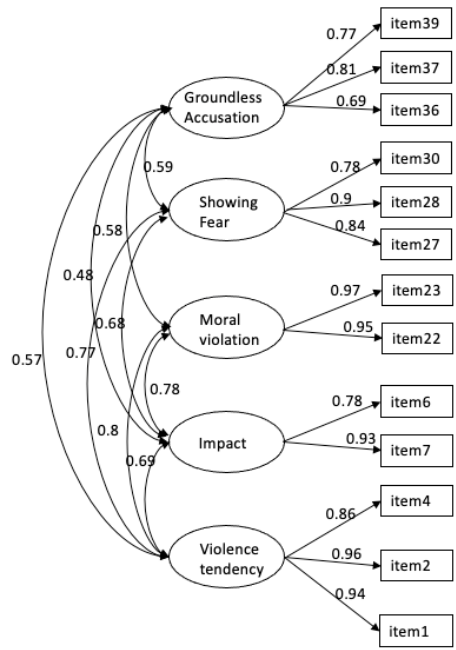

Figure 1 Five-factor model

Results from the exploratory factor analysis and confirmatory analysis thus provided initial evidence about the dimensionality of hate speech. Five distinctive factors were modeled adequately using 13 items. Three of the correlations between latent variables were in excess of r=.75 which would typically be a concern for discriminant validity. In this case, however, because only two stimuli have been used to generate variance in item responses, it is too early in the research process to combine or eliminate constructs on the basis of high inter-construct correlations. We plan to expand the amount of rated content in a future study, and with the addition of more examples of hate speech, a clearer picture of the inter-construct correlations will emerge.

### 4.3 Measurement Invariance Based on Age Differences

When collecting data, participants report their age based on five age groups, but n=143 participants were in the age group 18-33, so we run the multigroup analysis with two groups: older than 33 and younger than 33. We fitted the five-factor latent variable models and the subscales with different constraints. Our results showed that when constraining the error variance between age groups, the chi-squared differences were significant, which means that the two age groups do not exhibit measurement invariance. Table 4 shows this result. Tables 5 and 6 are follow-up analyses showing that Violence Tendency and Groundless Accusation were the sources of the invariance.

Table 4 Chi-Squared difference test of different age groups on the five-factor model

|  | DF | AIC | BIC | Chisq | Chisq diff | DF diff | Pr(>Chisq) |
|---|---|---|---|---|---|---|---|
| Configural | 110 | 5987.5 | 6294.8 | 155.52 |  |  |  |
| Weak | 118 | 5980.3 | 6262.5 | 164.32 | 8.806 | 8 | 0.3589 |

|        | DF  | AIC    | BIC    | Chisq  | Chisq diff | DF diff | Pr(>Chisq)      |
| ------ | --- | ------ | ------ | ------ | ---------- | ------- | --------------- |
| Strong | 126 | 5968.7 | 6225.9 | 168.73 | 4.410      | 8       | 0.8184          |
| Strict | 139 | 5986.7 | 6203.1 | 212.73 | 43.993     | 13      | 3.078e-05 ***   |

Signif. codes:  0 '***' 0.001 '**' 0.01 '*' 0.05 '.' 0.1 ' ' 1

Table 5 Chi-Squared difference test of different age groups on Violence Tendency

|            | DF | AIC    | BIC    | Chisq   | Chisq diff | DF diff | Pr(>Chisq)      |
| ---------- | -- | ------ | ------ | ------- | ---------- | ------- | --------------- |
| Configural | 0  | 1458.8 | 1515.3 | 0.0000  |            |         |                 |
| Weak       | 2  | 1459.7 | 1509.8 | 4.8633  | 4.8633     | 2       | 0.0878904       |
| Strong     | 4  | 1456.1 | 1500.0 | 5.2587  | 0.3954     | 2       | 0.8206158       |
| Strict     | 7  | 1469.9 | 1504.4 | 25.0856 | 19.8269    | 3       | 0.0001844 ***   |

Signif. codes:  0 '***' 0.001 '**' 0.01 '*' 0.05 '.' 0.1 ' ' 1

Table 6 Chi-Squared difference test of different age groups on the Groundless Accusation scale

|            | DF | AIC    | BIC    | Chisq   | Chisq diff | DF diff | Pr(>Chisq) |
| ---------- | -- | ------ | ------ | ------- | ---------- | ------- | ---------- |
| Configural | 0  | 1480.5 | 1537.0 | 0.0000  |            |         |            |
| Weak       | 2  | 1476.7 | 1526.9 | 0.1513  | 0.1513     | 2       | 0.92716    |
| Strong     | 4  | 1475.1 | 1519.0 | 2.6015  | 2.4503     | 2       | 0.29371    |
| Strict     | 7  | 1479.7 | 1514.2 | 13.1349 | 10.5333    | 3       | 0.01454 *  |

Signif. codes:  0 '***' 0.001 '**' 0.01 '*' 0.05 '.' 0.1 ' ' 1

### 4.4 Measurement Invariance Based on Personality Profiles

We used median splits to group personality profiles using the same logic as described above for age groups. The scores of each dimension are divided into high and low groups according to the comparison with the median. Our results showed that constraining the error variance between low and high groups, the chi-squared differences are significant on the five-factor multidimensional model. Besides performing significant differently on the five-factor model, high and low Extraversion significantly changed the evaluation of hate speech on Showing Fear (p=.01811) and Groundless Accusation (p=.017). High and low Agreeableness also influenced participants' evaluations on Groundless Accusation (p =.042) and Showing Fear (p =.018). High and low Conscientiousness significantly influenced the evaluation of hate speech on Showing Fear (p <.01) and Violence Tendency (p<.01). High and low Neuroticism significantly influenced the evaluation of Violence Tendency (p<.01) and Showing Fear (p <.01). High and low Openness to Experience significantly influenced the evaluation of hate speech on Violence tendency (p<.01). The significant *p*-value implies that the free and constrained models are significantly different. Some paths vary for different age groups and personality groups. In the other words, when evaluating hate speech, the aforementioned factors do not exhibit measurement invariance for different age groups and personality groups. For example, when constraining loadings and intercepts, different levels of Conscientiousness have significant influence on people's evaluation of hate speech related to violence. Table 7 shows the Chi-Squared different test of low and high Conscientiousness on the subscale Violence tendency.

Table 7 Chi-Squared difference test of low and high Conscientiousness on Violence tendency

|            | DF | AIC    | BIC    | Chisq   | Chisq diff | DF diff | Pr(>Chisq)   |
| ---------- | -- | ------ | ------ | ------- | ---------- | ------- | ------------ |
| Configural | 0  | 1414.2 | 1470.6 | 0.0000  |            |         |              |
| Weak       | 2  | 1410.9 | 1461.1 | 0.7420  | 0.7420     | 2       | 0.690056     |
| Strong     | 4  | 1413.0 | 1456.9 | 6.8724  | 6.1304     | 2       | 0.046645 *   |
| Strict     | 7  | 1420.7 | 1455.2 | 20.5401 | 13.6677    | 3       | 0.003394 **  |

Signif. codes:  0 '***' 0.001 '**' 0.01 '*' 0.05 '.' 0.1 ' ' 1

## 5 Discussion

### 5.1 Annotator's Subjectiveness in Hate Speech Annotation

Our results showed that age and personality differences were connected with the dimensional evaluation of hate speech. We used multi-group confirmatory factor analysis to assess measurement invariance across older and younger respondents, as well as across respondents who had high or low standing on each of the big five personality characteristics. Measurement invariance on self-report items examines differences in the ways that individuals respond to particular stimuli. In our study, respondents rated a particular content stimulus with a variety of items. Our significant results on strict invariance indicated that people had different patterns of responding to the items based on age and personality factors. Although these patterns do not directly tell us how their beliefs about hate speech may differ, we can strongly conclude that annotators' individual differences have a systematic relationship with variance in item responses to items designed to assess dimensional aspects of hate speech.

These results suggest that subjectivity plays a critical role in hate speech annotation. Older adult respondents, by dint of previous experience or knowledge that they possess relative to younger respondents, express different beliefs about content that advocates violence or that makes groundless accusations. Likewise, personality factors such as extraversion and agreeableness impact the patterns of item responses for content with respect to ratings of groundless accusations and fear towards a specific target population (such as a religious minority).

### 5.2 Design Implications

A key implication from these findings is that, while training annotators may provide improvements to inter-annotator agreement values, these improvements come at the expense of meaningful differences in how different people interpret different examples of speech. As Potter and Levine-Donnerstein [42] suggested, essential elements in projective latent content such as hate speech are symbols whose evaluation depends upon accessing pre-existing mental schemata. Those schemata vary across the lifespan and are based on naturally occurring individual differences, also implying that inter-annotator agreement of hate speech will only be strong to the extent that the annotators themselves are highly homogeneous.

These results also provide design insights for content moderation tools: Some evaluative dimensions that exhibit lesser degrees of inter-annotator variability could be construed as being more universal, and thus easier for an automated system to evaluate. On the other hand, evaluative dimensions with greater inter-rater variability would require a system or method that was adaptable to differing content moderation situations. If we assume that these latter dimensions cannot be consistently annotated (at least without an annotator training process that forces annotators all into the same "box"), then we will need to develop a more sophisticated understanding of the sources of variation in order to be able to develop a satisfactory annotation system.

Additionally, in manual content moderation, a community's moderators can decide the destiny of both posts and users in their community by promoting or hiding posts, as well as honoring, shaming, or banning the users. Moderators' decisions influence the content delivered to community members and audiences [21] and by extension also influence the community's experience of the discussion. Assuming that a human moderator is a community member who has demographic homogeneity with other community members, it seems possible that the mental schema they use to evaluate content will match those of other community members.

In contrast, both hybrid and fully-automated content moderation systems depend on trained machine learning models – the data for training these models comes from annotation processes [21]. These systems play an increasingly important role in content regulation [33]. For example, for identifying hate speech [10], pornography [51], and eating disorder-related content [9]. Given the apparent subjectivity of evaluations of certain dimensions of hate speech, hybrid and fully automated content moderation systems will need to either learn or be pre-programmed with an understanding of how speech evaluations may differ across demographic groups or communities. Further research will be needed to increase understanding of which evaluative dimensions generate the greatest variance in annotation results and how demographic differences are connected to specific patterns of evaluations. Conceivably, a machine learning system that is trained on not just on mean evaluations of speech examples, but also on degrees of variation, and demographic characteristics of annotators that are sources of variance, could then serve as the core of a content moderation system that was adaptable to different audiences or community contexts.

## 6 Limitations and Future Work

The present research was limited by the use of newly developed evaluative scales and by the use of the minimum number of speech instances needed to obtain variation in dimensional evaluations. Further validation work should be conducted to retest the items developed for this study. Additionally, the items should be tested on a wider variety of speech content, preferably using a method that includes multiple evaluations of each speech instance in order to partition evaluator variance from other sources of variance. In this study we use hate speech as an example. Our multi-dimensional scale approach is also applicable in other subjective data labeling tasks. For example, labeling misinformation and arthritics. We will explore these labeling tasks in the future.

## 7 Conclusion

In this paper we developed a multidimensional evaluative scale for hate speech and conducted several multigroup analysis to uncover the origin and value of disagreement among hate speech labelers. Our findings shown that personality and age had a substantial influence on the dimensional labeling of hate speech. These findings suggest that efforts to obtain annotation consistency among labelers with different backgrounds and personalities for hate speech may never fully succeed. Implications for the use of labeled data in automated or hybrid content moderation systems were discussed. Future research should be conducted to obtain additional validation evidence on these scales and to generalize to a wider variety of speech instances.